\documentclass[epj]{svjour}

\usepackage{graphics}
\usepackage{color}
\usepackage{booktabs}
\usepackage{flafter}
\usepackage{amsmath}

\begin{document}
\title{Systematic study of cross section for proton-induced reactions on Neodymium up to 65 MeV using TALYS-1.96 code}

\author{A.~Saha\inst{1} 
\thanks{\emph{email id:} arunabhaiitb@gmail.com}%
}                     
\offprints{}          
\institute{Department of Physics, ICFAI University Tripura, Kamalghat, Tripura 799210, India}
\date{17-September-2024}
\abstract{
{
Systematic study of theoretical estimation of nuclear cross section of charged particle induced reactions on rare earth nuclei has been carried out. The production cross section of the $^{150,149,148,146,144,143,141}$Pm and $^{149,147}$Nd nuclei were calculated theoretically via proton induced reaction on neodymium using TALYS (version 1.96) code in the default mode, with different combination of nuclear models as well as using adjusted nuclear model parameters from reaction threshold to the proton energy upto 65 MeV. The theoretically computed results were compared with the experimental results taken from EXFOR database and literature, reported by various research groups. Moreover, the effect of the various Level Density Models (LDMs), Preequilibrium Models (PEs), Optical Model Potentials (OMPs), gamma Strength Functions ($\gamma$SFs) in the cross section calculation were taken into consideration. The present theoretical analysis will help to understand the theory of nuclear reaction models and improve the evaluated nuclear data libraries.
}
} 
\authorrunning{A. Saha}
\titlerunning {Theoretical Calculations of ...}

\maketitle
\section{Introduction}      
\label{intro}

{ The estimation of cross section of charged particle induced reactions is very important for the design of an Accelerator-Driven System (ADS), the production of radioisotopes and for verfication of various nuclear models~\cite{yang2015,abyad2009,kim2014}. It is also important for understanding the origin of $p$-nuclei at astrophysical conditions~\cite{cheng2021}. Natural neodymium (Nd) is composed of 7 long-lived stable isotopes such as $^{142}$Nd (27.2\%), $^{143}$Nd (12.2\%),$^{144}$Nd (23.8\%), $^{145}$Nd (8.3\%), $^{146}$Nd (17.2\%), $^{148}$Nd (5.7\%) and $^{150}$Nd (5.6\%). Nd can be used as a target for producing radioisotopes such as $^{149}$Pm and $^{140,141}$Nd which are medically important and usable for therapeutic purposes~\cite{beyer2000,qaim2001,hil2005,tar2014}. 

Knowledge of production cross section of proton induced reactions on natural Nd is also important for the experimental observation of doube $\beta$ decay. $^{150}$Nd, which is one of the 7 long naturally occuring isotopies of $^{nat}$Nd, is a very promising candidate for the observation of neutrinoless double $\beta$ decay ($\beta\beta0\nu$) having a very large Q-value of 3371.38 $\pm$0.20 keV~\cite{Kol10}. Since the expected signal strength of double $\beta$ decay of $^{150}$Nd is extremely weak, reduction in background is very much required. There are, at present, three ongoing experiments where neutrinoless double $\beta$ decay of $^{150}$Nd is planned to be studied. These are Sudbury Neutrino Observatory plus liquid scintillator (SNO+)~\cite{Man11}, Drift Chamber Beta-ray Anallyzer (DCBA)~\cite{Ish10} and SuperNEMO~\cite{Arn10}. One of the background component in the double $\beta$ decay experiment is the decay of long-lived radioisotopes viz., $^{143}$Pm, $^{144}$Pm, $^{146}$Pm, $^{148}$Pm, $^{147}$Nd etc. produced in natural Nd by its reaction with the proton from cosmic ray. It, therefore, necessitates the measurement of excitation functions of proton induced reactions on natural Nd.

The goal of this paper is to carry out systematic study of theoretical estimation of production cross section of various medically important isotopes of lanthanides via proton induced reaction on neodymium and to compare the calculated cross sections with the experimental results obtained (using natural and enriched Nd targets) till date during various experiments. Such systematic studies are important for validating the theoretical calculations done by TALYS code (version 1.96)~\cite{TALYS}, taking into account the various Optical Model Potentials (OMPs), Level Density Models (LDMs) and Preequilibrium models (PEs).  
}

\section{Details of the experimental data}
\label{sec:1}

{

For this work, the experimental data has been taken from the work of J.~Olkowsky {\em et al.}~\cite{Olk61,EXFOR}, O.~Lebeda {\em et al.}~\cite{Leb12,Leb14}, D. Banerjee {\em et al.}~\cite{Bhatt15}, S.~C.~Yang {\em et al.}~\cite{yang2015} and F.~Tarkanyi {\em et al.}~\cite{Tarkanyi2017}. In the work of J.~Olkowsky~{\em et al.}, the cross sections of the $^{149,150}$Pm produced via. proton-induced reactions on $^{150}$Nd have been measured. The chosen energy range of the incident proton beam was from 6.9 to 10.84 MeV. In the work of O.~Lebeda {\em et al.}~\cite{Leb12,Leb14}, proton-induced reactions on {\em natural} neodymium ($^{nat}$Nd) was carried out in the 5-35 MeV energy range in order to measure the production cross sections of different radionuclides. Similarly, in the work of D.~Banerjee~{\em et al.}~\cite{Bhatt15}, the excitation function of the $^{150}$Nd($p$,x$n$y$p$) reactions have been measured using the stacked-foil activation technique with a 97.65\% {\em enriched} $^{150}$Nd target. The range of proton energy chosen was 7-15 MeV. In that reported work~\cite{Bhatt15}, we had carefully measured the absolute Cross Sections (CSs) for $^{150}$Nd($p$,$n$)$^{150}$Pm, $^{150}$Nd($p$,2$n$)$^{149}$Pm and $^{150}$Nd($p$;$d$,$pn$)\\$^{149}$Nd reactions at various proton energy values ranging from 7-15 MeV for the first time. It is worth emphasizing here that the dead time of the detectors used in the work of D.~Banerjee {\em et al.}~\cite{Bhatt15} was significantly lower ($\le$10$\%$) than that allowed in the work of O.~Lebeda {\em et al.}~\cite{Leb12,Leb14}, i.e., ($\le$40$\%$). In the work of S.~C.~Yang {\em et al.}~\cite{yang2015}, the production cross sections of $^{nat}$Nd($p$,x) reactions were determined by using stacked-foil activation technique for proton energy upto 45 MeV. In the work of  F.~Tarkanyi {\em et al.}~\cite{Tarkanyi2017}, the excitation functions of the $^{nat}$Nd($p$,x) nuclear reactions were measured upto 65 MeV energy, using stacked foil activation technique and high resolution $\gamma$-spectrometry.

}

\section{Theoretical calculations}
\label{sec:1}
In this work, systematic study of the theoretical estimations of cross sections of $^{150,149,148,146,144,143,141}$Pm and $^{149,147}$Nd nuclei were carried out via proton-induced reaction on neodymium target using the TALYS code for incident proton beam with energies upto 65 MeV. TALYS is a theoretical nuclear model code in which photons, neutrons, protons, deuterons, $^3$He and $^4$He can be used as projectiles in the energy range of 1 keV to 200 MeV for target elements with mass of 12 and heavier~\cite{TALYS}. It utilizes state-of-the-art nuclear reactions models and includes direct reaction, preequilibrium emission and compound reaction mechanisms. The various experimental results taken from EXFOR~\cite{EXFOR} and literature were compared with the theoretically calculated values based on the TALYS-1.96 code. In the present work, the theoretical calculations for each individual reaction were at first carried out using various level density models (LDMs) in the default mode and then using various combinations of nuclear model parameters to fine tune the estimated cross sections with the observed experimental results. Finally, the calculations were done using adjusted global parameters for the purpose of fine tuning. The level density parameters were calculated using six different choices of the level density models available in TALYS-1.96. In TALYS, the six different nuclear level density models used for calculating the nuclear reaction cross section are as follows:-
(i) LDM-1: Constant temparature model~\cite{Gilbert1965} + Fermi Gas model~\cite{Dilg1973} (default)
(ii) LDM-2: Back-shifted Fermi gas model~\cite{Dilg1973} 
(iii) LDM-3: Generalized Superfluid model~\cite{Ignatyuk1979,Ignatyuk1993} 
(iv) LDM-4: Microscopic model of Goriely~\cite{Goriely2001} based on Hartree-Fock calculations
(v) LDM-5: Microscopic Combinatorial model of S.~Goriely-S.~Hilaire~\cite{Goriely2008} 
(vi) LDM-6: Microscopic level density based on temparature dependent Hartree-Fock-Bogoliubov calculations~\cite{Hilaire2012} using the D1M Gogny force. 

In the Fermi gas model, the level density parameter 'a' can be derived from D$_{0}$ which is the s-wave average neutron resonance spacing (in keV) at the neutron separation energy. In the work, the experimental value of D$_{0}$ for the $^{150}$Nd was obtained from the RIPL-3 database~\cite{Capote2009} and compared with the theoretical values predicted  by the TALYS code for each level density model. The values of the D$_{0}$ are mentioned in Table~\ref{tab:1}.

\begin{table}
\caption{The experimental value of D$_{0}$ for the $^{150}$Nd was obtained from the RIPL-3 database and compared with the theoretical values predicted for each level density model by the TALYS-1.96 code}
\label{tab:1}       
\begin{tabular}{lll}
\hline\noalign{\smallskip}
Level density model & Experimental & Theoretical \\
		    & value of D$_{0}$& value of D$_{0}$ \\
                    & (eV)           & (eV)  \\
\noalign{\smallskip}\hline\noalign{\smallskip}
Constant temparature  & 0.165$\pm$0.015 & 6.21 \\
Back-shifted Fermi gas &  & 2.78\\
Generalized superfluid &  & 3.47\\
Goriely &  & 11.13\\
Goriely-Hilaire&  & 2.94\\
Goriely-Hilaire Gogny force&  & 2.39\\
\noalign{\smallskip}\hline
\end{tabular}

\end{table}

In TALYS, the calculation of the contribution of compound nucleus is done using the Hauser-Feshbach model~\cite{Hauser1952}. The preequilibrium contribution was calculated using the exciton model~\cite{Kalbach1986} and the multistep direct/compound model. The optical model potential (OMP) by Koning-Delaroche (KD)~\cite{Koning2004} was used. In order to consider the $\gamma$-ray emission channel, TALYS uses a fixed $\gamma$-ray strength function ($\gamma$SF) model. In the present work, the Brink-Axel (BA) Lorentzian~\cite{Brink1957,Axel1962} and the Kopecky-Uhl (KU) generalized Lorentzian~\cite{Kopecky1990} were used as the  $\gamma$SFs. 

In this work, combination of various input parameters and models were used assuming the fact that the cross sections of nuclear reactions are very much dependent on the nuclear level densities (NLDs), optical model potentials (OMPs) and $\gamma$-ray strength functions ($\gamma$ SFs). 

\section{Results and Discussion}

\subsection{Comparison of various experimental data and TALYS evaluated Nuclear Data Library (TENDL)}

\begin{figure}

\resizebox{0.5\textwidth}{!}{%
\includegraphics{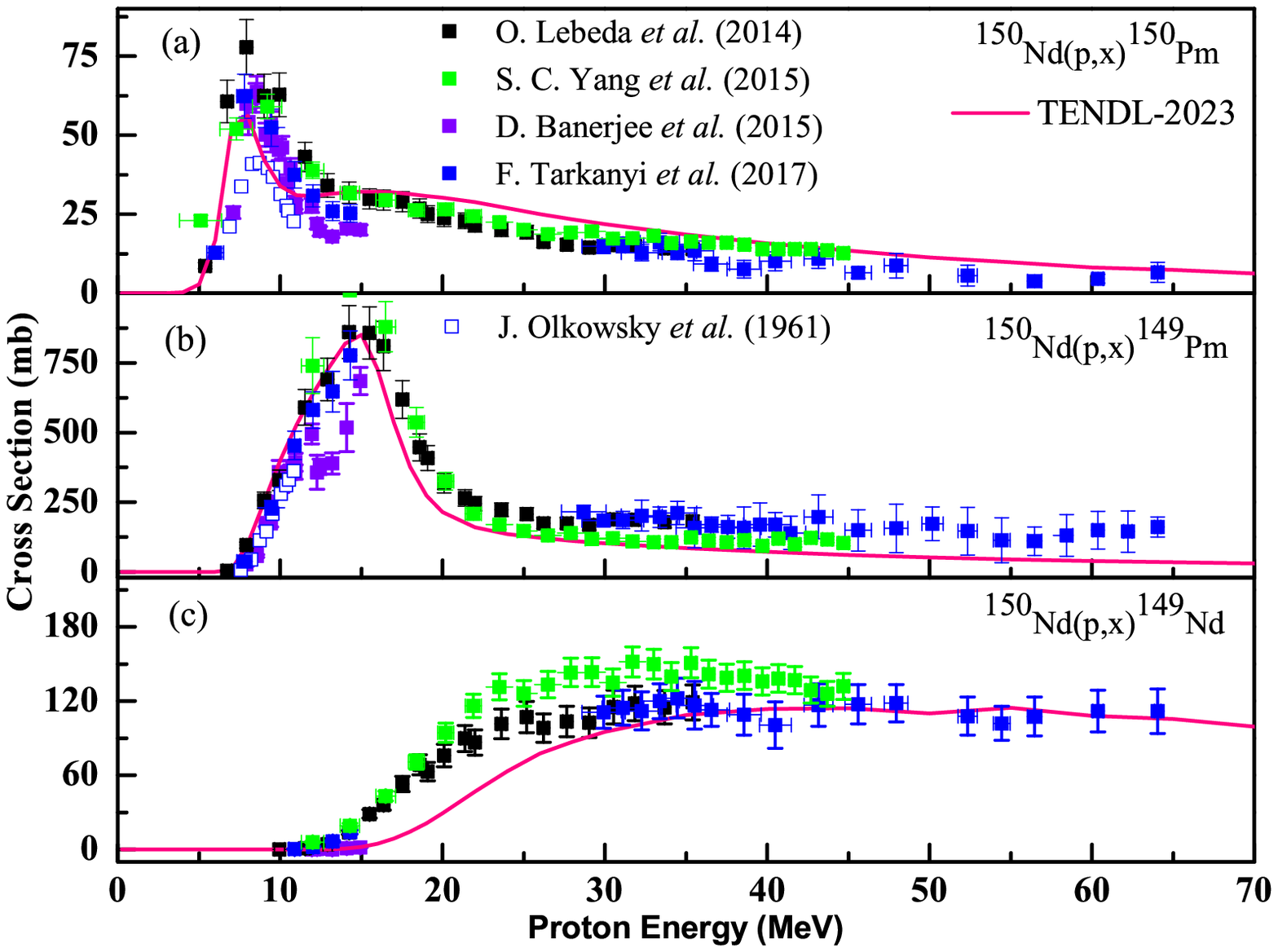}
}

\caption{The data of D.~Banerjee {\em et al.}~\cite{Bhatt15} for the (a) $^{150}$Nd($p$,$n$)$^{150}$Pm, (b) $^{150}$Nd($p$,2$n$)$^{149}$Pm and (c) $^{150}$Nd($p$;$d$,$pn$)$^{149}$Nd reactions compared with the previously published work of J. Olkowsky {\em et al.}~\cite{Olk61} taken from EXFOR database~\cite{EXFOR}, S.~C.~Yang {\em et al.}~\cite{yang2015}, F.~Tarkanyi {\em et al.}~\cite{Tarkanyi2017} and  O.~Lebeda {\em et al.}~\cite{Leb12,Leb14} together with the TENDL-2023 data~\cite{Koning2019}.}
\label{fig1}       
\end{figure}

\begin{figure*}[!]

\resizebox{\textwidth}{!}{%
\includegraphics{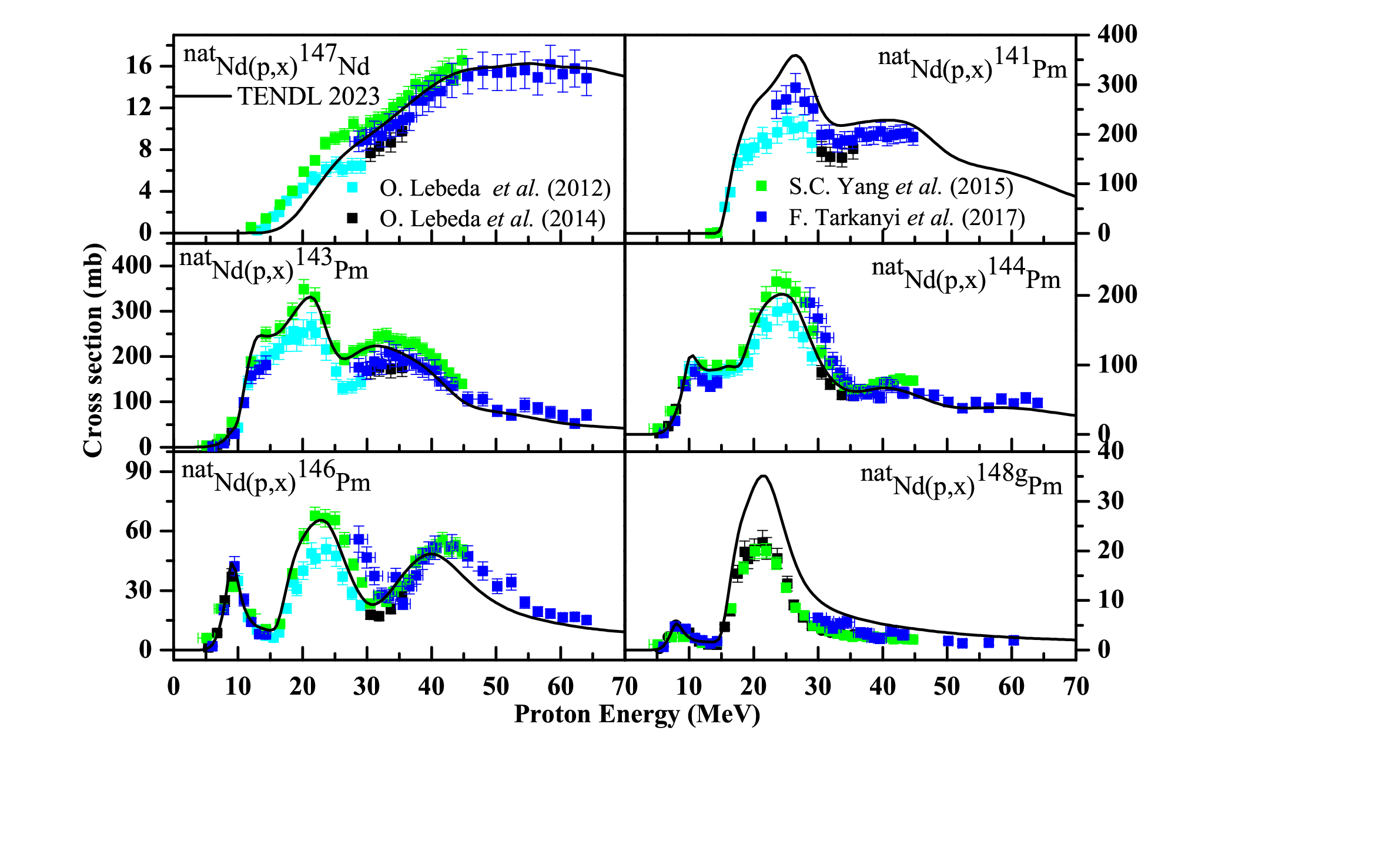}
}

\caption{Experimental data for $^{nat}$Nd($p$,x)$^{148g,146,144,143,141}$Pm,$^{147}$Nd taken from EXFOR database~\cite{EXFOR} together with the TENDL-2023 data~\cite{Koning2019}.} 
\label{fig2}       
\end{figure*}

The results on production cross sections of $^{150,149}$Pm and $^{149}$Nd reported by D.~Banerjee {\em et al.}~\cite{Bhatt15} are shown in panel (a), (b) and (c) of Fig.~\ref{fig1} along with the literature data of O.~Lebeda {\em et al.}~\cite{Leb12,Leb14} and J.~Olkowsky {\em et al.} taken from the EXFOR compilation~\cite{EXFOR} and compared with the evaluated data from TENDL-2023~\cite{Koning2019} database. This is to mention here that the CS values, obtained by the work of O.~Lebeda {\em et al.} via bombardment of proton on natural $^{150}$Nd target, were scaled up to the absolute values obtained by D.~Banerjee {\em et al.}~\cite{Bhatt15} taking the isotopic enrichment of $^{150}$Nd into consideration. This is because in the work of D. Banerjee {\em et al.}~\cite{Bhatt15}, enriched $^{150}$Nd target was used. Similarly, in Fig.~\ref{fig2}, the measured cross section data of $^{148,146,144,143,141}$Pm and $^{147}$Nd taken from  EXFOR compilation~\cite{EXFOR} and produced experimentally via proton-induced reaction on $^{nat}$Nd by O.~Lebeda {\em et al.}~\cite{Leb12,Leb14}, S.~C.~Yang {\em et al.}~\cite{yang2015} and F.~Tarkanyi {\em et al.}~\cite{Tarkanyi2017} have been shown and compared with the evaluated data from TENDL-2023~\cite{Koning2019} database. 

\subsection{Comparison of various experimental data with the theoretical values based on the TALYS code using default mode as well as combination of model parameters}

The TALYS calculations were carried out for the $^{150}$Nd($p$,x)\\$^{150,149}$Pm,$^{149}$Nd and $^{nat}$Nd($p$,x)$^{148,146,144,143,141}$Pm, $^{147}$Nd reactions using the default parameters adopting the different  phenomenological and microscopic level density models together with Koning-Delaroche (KD) local optical model potentials and Simplified Model Lorentzian (SMLO) E1 $\gamma$-ray strength function ($\gamma$SF). Along with the default mode, the theoretical calculations were also carried out using the combination of various model parameters for each of the above mentioned reactions in order to fine tune the theoretical results with the experimental data. The results of the theoretical calculations (both using the default mode as well as using different combination of model parameters) for the various reactions are shown in Fig.~\ref{fig3} to Fig.~\ref{fig11}.

\subsubsection{$^{150}$Nd($p$,x)$^{150}$Pm reaction}

\begin{table*}
\begin{center}

\rotatebox{90}{
\label{tab2}
\begin{tabular}{|| c c c c c c||} 
 \hline
  Level Density Model& Optical model Potential&$\gamma$SFs&Preequilibrium model& widthmode& parameters\\ [0.5ex] 
 \hline\hline
 Constant Temparature& Koning-Delaroche & Brink-Axel & preeqmode 3 & HRTW model & fullhf y \\ 
&local & Lorentzian&&&asys y\\
& & &&&strength 2\\
& & &&&widthmode 2\\
 \hline
 Back-shifted& Koning-Delaroche &Brink-Axel &preeqmode 3&HRTW model& fullhf y \\
 Fermi gas&local dispersive&Lorentzian&&&asys y \\
& potential&&&&strength 2 \\
& & &&&widthmode 2\\
 \hline
Generalized &Koning-Delaroche&Brink-Axel &preeqmode 3&HRTW model&fullhf y\\
superfluid&global&Lorentzian&&&asys y\\
&&&&&gshell y\\
&&&&&strength 2\\
&&&&&widthmode 2\\
 \hline
Goriely&Koning-Delaroche&Kopecky-Uhl&preeqmode 3&HRTW model&fullhf y\\
&global&generalized&&&gshell y\\
&&lorentzian&&&strength 1\\
&&&&&widthmode 2\\
\hline
Goriely-Hilaire&Koning-Delaroche&Kopecky-Uhl&preeqmode 3&HRTW model&fullhf y\\
&global&generalized&&&strength 1\\
&&lorentzian&&&widthmode 2\\
 \hline
Goriely-Hilaire &Koning-Delaroche&Brink-Axel &preeqmode 3&HRTW model&fullhf y\\
Gogny force&global&Lorentzian&&&strength 2\\
&&&&&widthmode 2\\
\hline
\end{tabular}
}
\caption{Details of the different statistical models and parameters used in TALYS calculations for $^{150}$Nd($p$,x)$^{150}$Pm reaction}  
\end{center}
\end{table*}

\begin{figure}
\resizebox{0.5\textwidth}{!}{%
\includegraphics{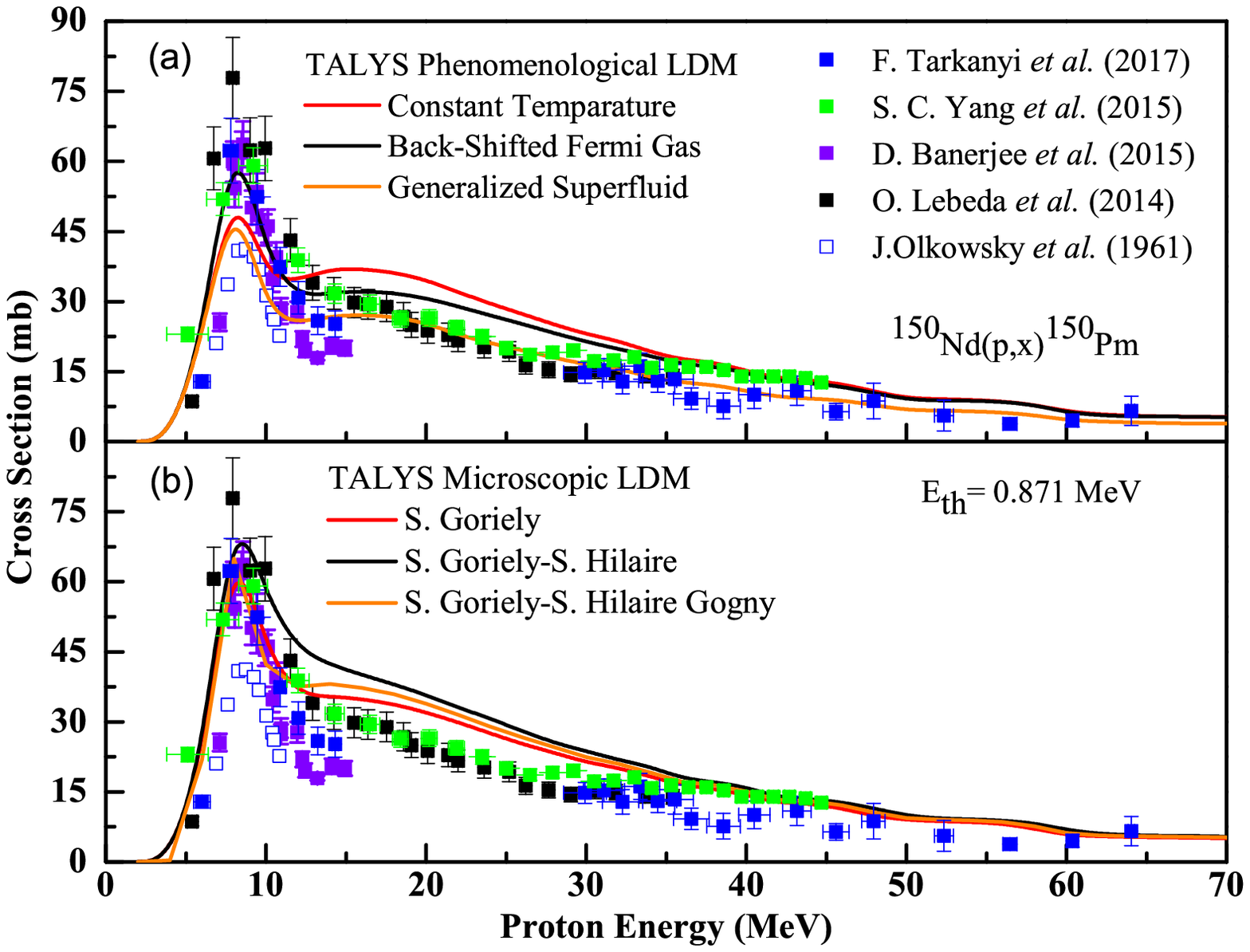}
}

\caption{The experimental data for the $^{150}$Nd($p$,x)$^{150}$Pm reaction from the previous works of S.~C.~Yang {\em et al.}~\cite{yang2015}, J.~Olkowsky {\em et al.}~\cite{Olk61}, O.~Lebeda {\em et al.}~\cite{Leb12,Leb14}, D.~Banerjee {\em et al.}~\cite{Bhatt15} and F.~Tarkanyi {\em et al.}~\cite{Tarkanyi2017} compared with the (a) phenomenological and (b) microscopic level density model calculations based on TALYS code with the default option. Discrepancies have been observed between the experimental data and the theoretically predicted cross section values at various energies calculated using the six different level density models in the default mode.}
\label{fig3}      
\end{figure}

For $^{150}$Nd($p$,x)$^{150}$Pm reaction (see Fig.~\ref{fig3}), the Constant Temparature model in the default mode could reproduce well the trend of variation of the cross section data reported by the various groups upto 10 MeV proton energy. Beyond this energy and upto 35 MeV, the theoretical results generated by the constant temparature model overestimate the experimental data reported by D.~Banerjee {\em et al.}~\cite{Bhatt15}, O.~Lebeda {\em et al.}~\cite{Leb12,Leb14}, S.~C.~Yang {\em et al.}~\cite{yang2015} and F.~Tarkanyi {\em et al.}~\cite{Tarkanyi2017}. From 35-45 MeV proton energy range, the theoretically calculated cross section by the constant temparature model matches well with the data reported by S. C. Yang {\em et al.}~\cite{yang2015}. Again from 45-65 MeV range, it matches closely with the data reported by F.~Tarkanyi {\em et al.}~\cite{Tarkanyi2017}. The Back Shifted Fermi Gas model in its default mode matches the experimental results reported by D.~Banerjee {\em et al.}~\cite{Bhatt15} only upto $\sim$12.5 MeV proton energy. It matches the trend of variation and magnitude of the results reported by O. Lebeda~{\em et al.}~\cite{Leb12,Leb14} reasonably well upto energy 17.5 MeV. Beyond that, discrepancy is observed between the theoretical and the experimental results upto 30 MeV. Beyond this energy, the CS prediction of this model matches well with the data reported by S. C. Yang {\em et al.}~\cite{yang2015} in the 35-45 MeV range and with the F.~Tarkanyi {\em et al.}~\cite{Tarkanyi2017} in the 45-65 MeV range. Between 11-15 MeV proton energy, discrepancy is observed between the theoretical calculations of the Generalized Superfluid model in the default mode and the results reported by D. Banerjee {\em et al.}~\cite{Bhatt15}. The shape and magnitude of the excitation function produced by this model matches well with the results reported by J. Olkowsky {\em et al.}~\cite{Olk61} in the 5-10 MeV range. From 15-35 MeV range, it matches well with that of O.~Lebeda {\em et al.} and S.~C.~Yang {\em et al.}~\cite{yang2015}. Beyond 35 MeV, the predictions of this model matches only with the data of F.~Tarkanyi {\em et al.}~\cite{Tarkanyi2017}. The default theoretical results of all the microscopic level density models (i.e., LDM-4, LDM-5 and LDM-6) (see panel-(b) of Fig.~\ref{fig3}) overestimates the experimental results reported by O. Lebeda {\em et al.}~\cite{Leb12,Leb14}, S.~C.~Yang {\em et al.}~\cite{yang2015} and D. Banerjee {\em et al.}~\cite{Bhatt15} beyond 11-12 MeV proton energy and upto 35 MeV. Beyond 35 MeV, it matches closely with the data reported by S.~C.~Yang {\em et al.}~\cite{yang2015} upto 45 MeV and with that of F.~Tarkanyi~{\em et al.}~\cite{Tarkanyi2017} in the 45-65 MeV range. The variation pattern of theoretical results for all the phenomenological and microscopic level density models is similar to that reported by J. Olkowsky {\em et al.} in the 6-11 MeV range.

\begin{figure*}[!]

\resizebox{\textwidth}{!}{%
  \includegraphics{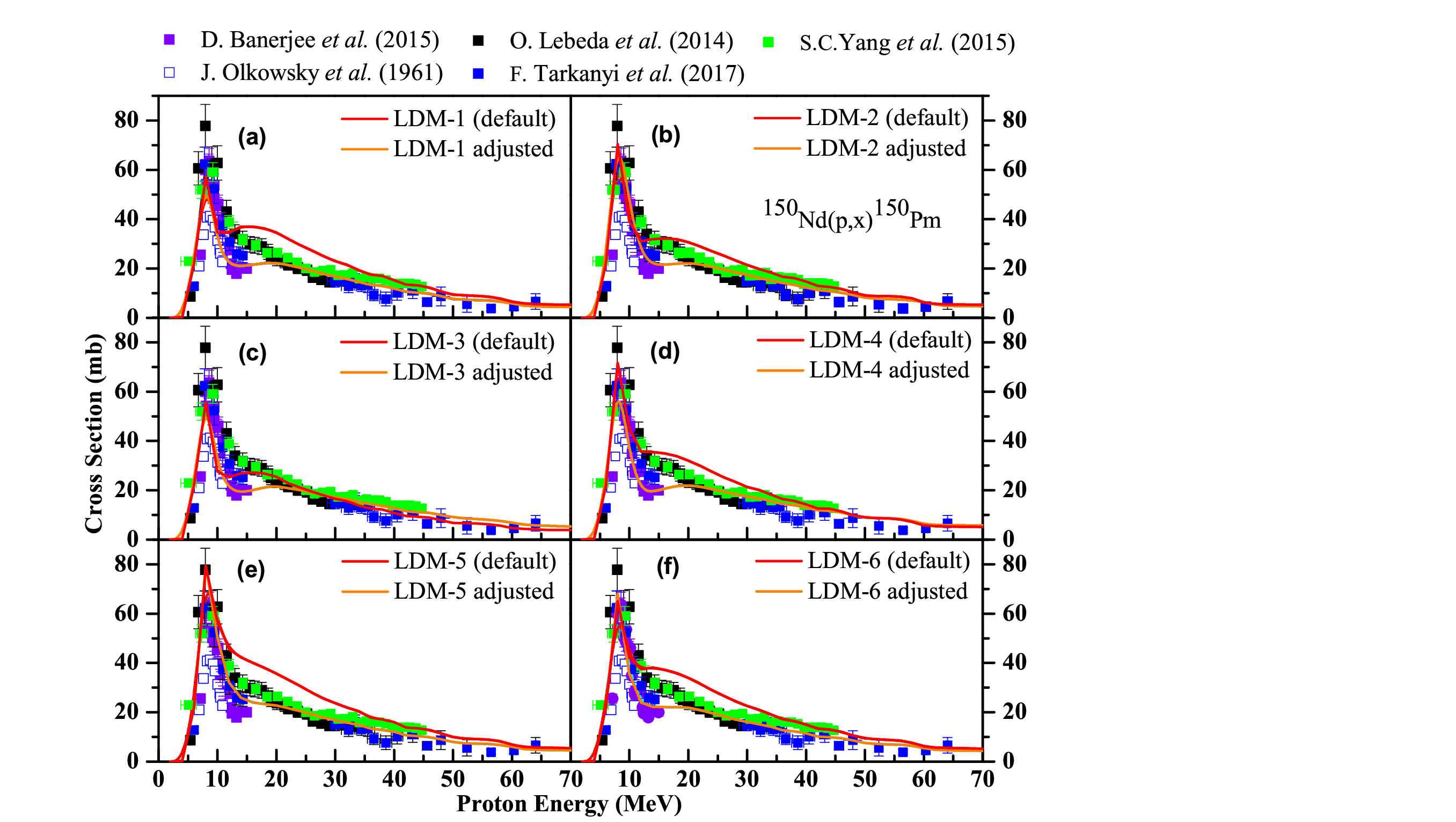}
}
\caption{The experimentally measured data for the $^{150}$Nd($p$,x)$^{150}$Pm reaction obtained from literature compared with the phenomenological theoretical calculations based on TALYS code with different combination of nuclear models and model parameters. The red and the sky-blue continuous lines indicate the theoretically estimated CSs using TALYS in the default mode and using different combination of model parameters, respectively}
\label{fig12}      
\end{figure*}

Keeping in view the large discrepancies between the measured cross section and the theoretical results from the TALYS using default parameters
, the theoretical values were revised with different combinations of nuclear models and/or model parameters in an attempt to reproduce the experimental results more accurately. While doing so, the optical model potential, level density models and preequilibrium models, which are of utmost importance, have been individually adjusted. For example, the behavior of the Constant Temparature model was significantly improved by combining the Koning-Delaroche (KD) local optical potential with Brink-Axel Lorentzian as the gamma strength function and preequmode 3 as the preequilibrium model. The response of the Back-Shifted fermi gas model was improved by combining KD local dispersive model with Brink-Axel Lorentzian and preeqmode 3. For both these aforesaid level density models, the parameters ''asys" and ''fullhf" were enabled to match with the experimental data. The parameter ''asys" is a flag to use all level density parameters from systematics by default and ''gshell'' is used to include the damping of shell effects with excitation energy in single-particle level densities. The Generalized Superfluid model was improved by combining KD global potential with Brink-Axel Lorentzian. For this model, the parameters "asys" and "gshell" were enabled. The response of the microscopic models of Goriely and Goriely-Hilaire was improved by the combination of KD global optical model potential and Kopecky-Uhl (KU) generalized Lorentzian $\gamma$SF along with preequilibrium model 3. For the former microscopic model of Goriely, the parameter "gshell" was enabled. Similarly, the behavior of the microscopic model of Goriely-Hilaire Gogny force was improved by combining KD global optical model potential with Brink-Axel Lorentzian and preequilibrium model 3. For all the aforementioned phenomenological and microscopic level density models, the parameter "fullhf" is enabled and the Hofmann-Richert-Tepel-Weidenm$\ddot{u}$ller (HRTW) model was used. This is worth mentioning here that the parameter "fullhf" takes care of j-l coupling in the Hauser-Feshback theory. Kindly see Fig.~\ref{fig12} and refer to Table~2 for more elaborate details.

\subsubsection{$^{150}$Nd($p$,x)$^{149}$Pm reaction}

\begin{figure*}[!]

\resizebox{\textwidth}{!}{%
  \includegraphics{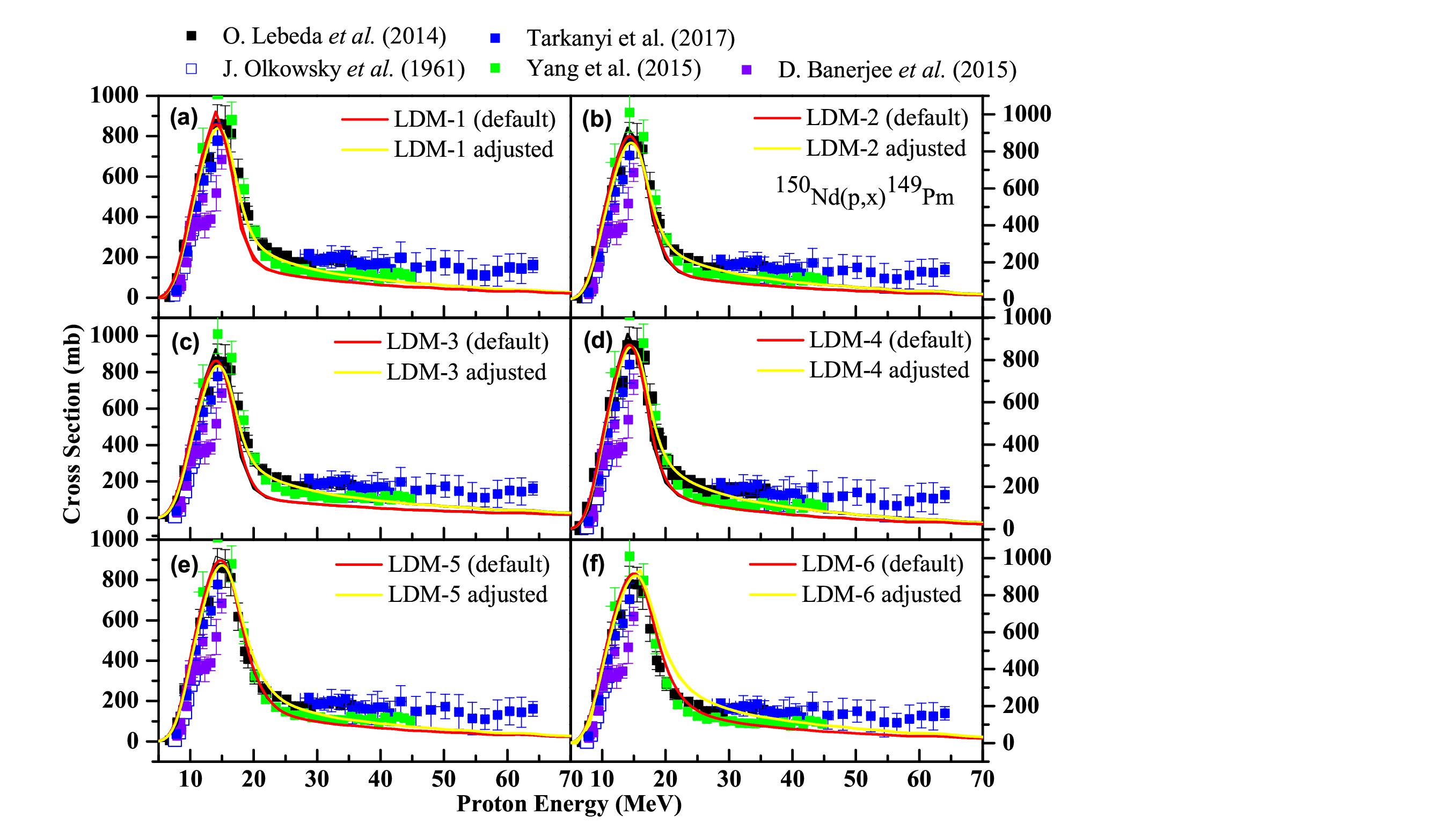}
}
\caption{The experimentally measured data for the $^{150}$Nd($p$,x)$^{149}$Pm reaction obtained from literature compared with the phenomenological theoretical calculations based on TALYS code with different combination of nuclear models and/or model parameters. The black and red continuous lines indicate the theoretically estimated CSs using the default mode and different combination of nuclear models, respectively.}
\label{fig13}      
\end{figure*}

Similarly, for $^{150}$Nd($p$,x)$^{149}$Pm reaction (see Fig.~\ref{fig13}), the excitation function generated by the theoretical calculations of all the phenomenological and microscopic models peaks at around 15 MeV. Upto this energy value, the theoretical results calculated by the various models matches with the experimentally calculated cross section values reasonably well (except the CS values at 12.44, 13.21, 14.10 and 14.94 MeV proton energy reported by D.~Banerjee {\em et al.}~\cite{Bhatt15}). Beyond the peak value, the phenomenological level density models underestimates the cross section values. Though the microscopic level density models produces results of similar trend as that of the experimental results, little overestimation and underestimation is observed at some energy values in the 15-65 MeV range.

However, in an attempt to improve the behavior of the Constant Temparature model, combination of KD local optical model potential and Brink-Axel Lorentzian were used together with enabled "asys" and "gshell" parameters. Similarly, for the Back-Shifted Fermi gas and the Generalized Superfluid models, the combinations of KD local dispersive model \& KU generalized Lorentzian and KD global \& Brink-Axel Lorentzian were used respectively as OMP and $\gamma$SF. The parameter "asys" was also enabled for these level density models with a view to improve their behavior. For the microscopic models of Goriely \& Goriely-Hilaire, the combination of KD global OMP and KU generalized lorentizian $\gamma$SF were used together with enabled parameters of "gshell" and/or "asys". For the level density model of Goriely-Hilaire Gogny force, the KD global OMP is combined with KU generalized Lorentzian. Also, for all the above mentioned LDMs, widthmode 2 (i.e., HRTW model) is used and "fullhf" parameter is enabled. The parameter ''widthmode'' is used to include model for width mode fluctuation corrections in compound nucleus calculations. For all the level density models, preequilibrium model 4, i.e.,  multi-step direct or compound model is used. Kindly refer to Fig.~\ref{fig13} and Table 3 for more details.

\begin{table*}
\begin{center}

\rotatebox{90}{
\label{table-3}
\begin{tabular}{|| c c c c c c||} 
 \hline
  Level Density Model& Optical model Potential&$\gamma$SFs&Preequilibrium model& widthmode& parameters\\ [0.5ex] 
 \hline\hline
 Constant Temparature& Koning-Delaroche local & Brink-Axel Lorentzian& preeqmode 4 & HRTW model & fullhf y \\ 
& & &&&strength 2\\
& & &&&widthmode 2\\
& & &&&asys y\\
& & &&&gshell y\\
 \hline
 Back-shifted Fermi gas& Koning-Delaroche local&Kopecky-Uhl generalized &preeqmode 4&HRTW model& strength 1 \\
 & dispersive Lorentzian&Lorentzian&&&fullhf y \\
& &&&&widthmode 2 \\
& & &&&asys y\\
 \hline
Generalized superfluid&Koning-Delaroche global&Brink-Axel Lorentzian&preeqmode 4&HRTW model&fullhf y\\
&&&&&asys y\\
&&&&&widthmode 2\\
 \hline
Goriely&Koning-Delaroche global&Kopecky-Uhl generalized&preeqmode 4&HRTW model&fullhf y\\
&& lorentzian&&&gshell y\\
&&&&&ptable 61 149 -0.25238\\
&&&&&widthmode 2\\
\hline
Goriely-Hilaire&Koning-Delaroche global&Kopecky-Uhl generalized&preeqmode 4&HRTW model&fullhf y\\
&& lorentzian&&&strength 1\\
&&&&&widthmode 2\\
&&&&&asys y\\
&&&&&gshell y\\
 \hline
Goriely-Hilaire Gogny&Koning-Delaroche global&Kopecky-Uhl generalized &preeqmode 4&HRTW model&fullhf y\\
 force&&lorentzian &&&strength 1\\
&&&&&widthmode 2\\
\hline
\end{tabular}
}
\caption{Details of the different statistical models and parameters used in TALYS calculations for $^{150}$Nd($p$,x)$^{149}$Pm reaction}  
\end{center}
\end{table*}

\subsubsection{$^{150}$Nd($p$,x)$^{149}$Nd reaction}

\begin{figure*}[!]

\resizebox{\textwidth}{!}{%
  \includegraphics{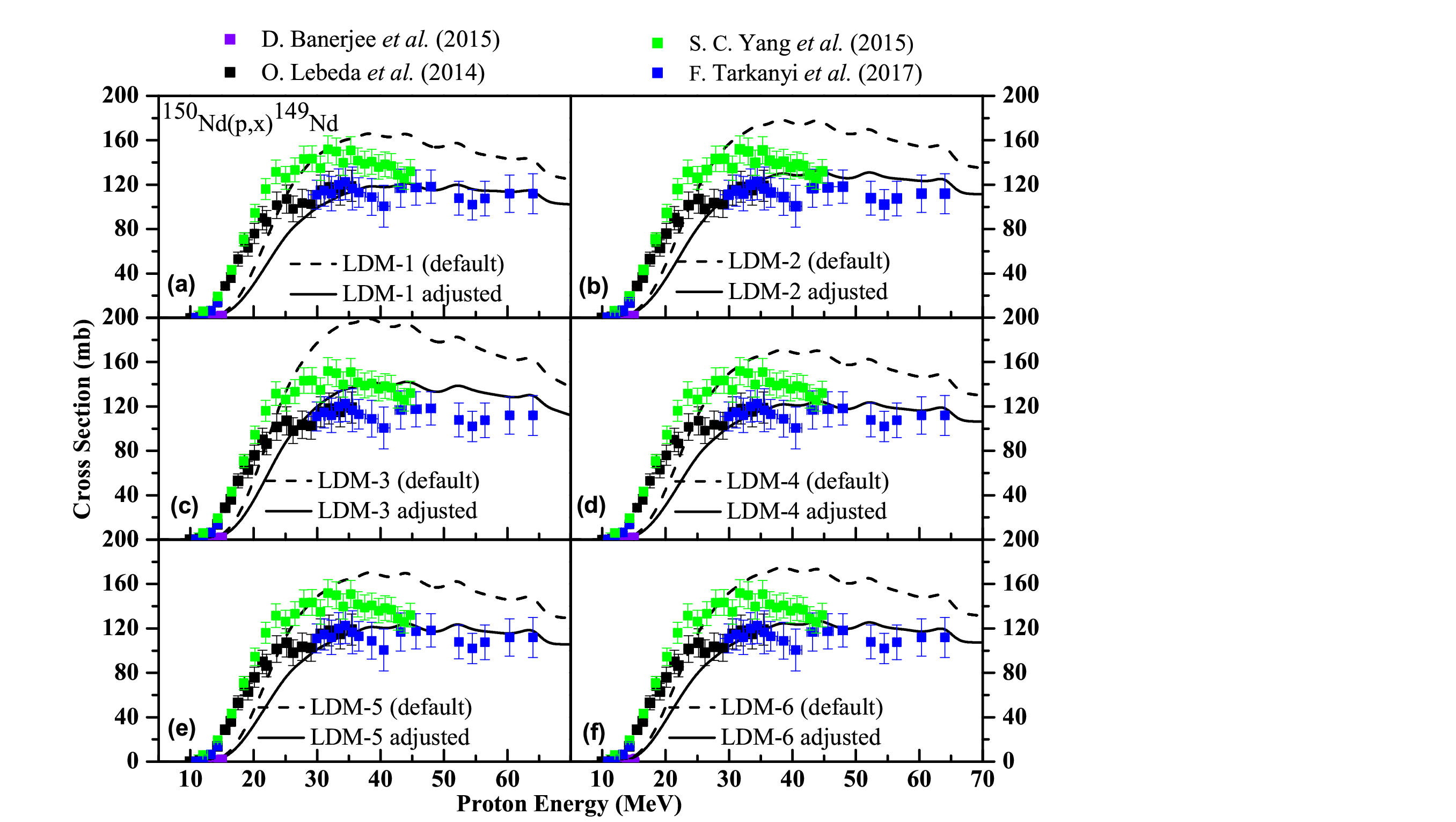}
}
\caption{The experimentally measured data for the $^{150}$Nd($p$,x)$^{149}$Nd reactions obtained from literature compared with the phenomenological theoretical calculations based on TALYS code in the default mode and with different combination of model parameters. The dotted and the continuous lines indicate the theoretically estimated CSs using TALYS in the default mode and using different combination of model parameters, respectively}
\label{fig17}       
\end{figure*}

For $^{150}$Nd($p$,x)$^{149}$Nd reaction (see Fig.~\ref{fig17}), the excitation functions generated by the theoretical calculations of all the level density models (both phenomenological and microscopic) follows the trend of variation and also the magnitude of experimental data reported by the various groups in a satisfactory manner upto 30 MeV. Beyond this energy, the theoretical calculations overestimates the cross section values and does not give satisfactory results in the default mode. 

However, the response of the various level density models ranging from LDM-1 to LDM-6 improves significantly for $^{150}$Nd($p$,x)$^{149}$Nd reactions when the preequilibrium model is changed from the preeqmode 2 (default exciton model) to the preeqmode 3 (numerical transition rates with optical model for collision probability). Kindly refer to Fig.~\ref{fig17} %
to see the effect of the change of the preequilibrium mode in the response of the different level density models for the respective reaction channels.

\subsubsection{$^{nat}$Nd($p$,x)$^{148g}$Pm reaction}

\begin{figure*}

\resizebox{0.7\textwidth}{!}{%
\includegraphics{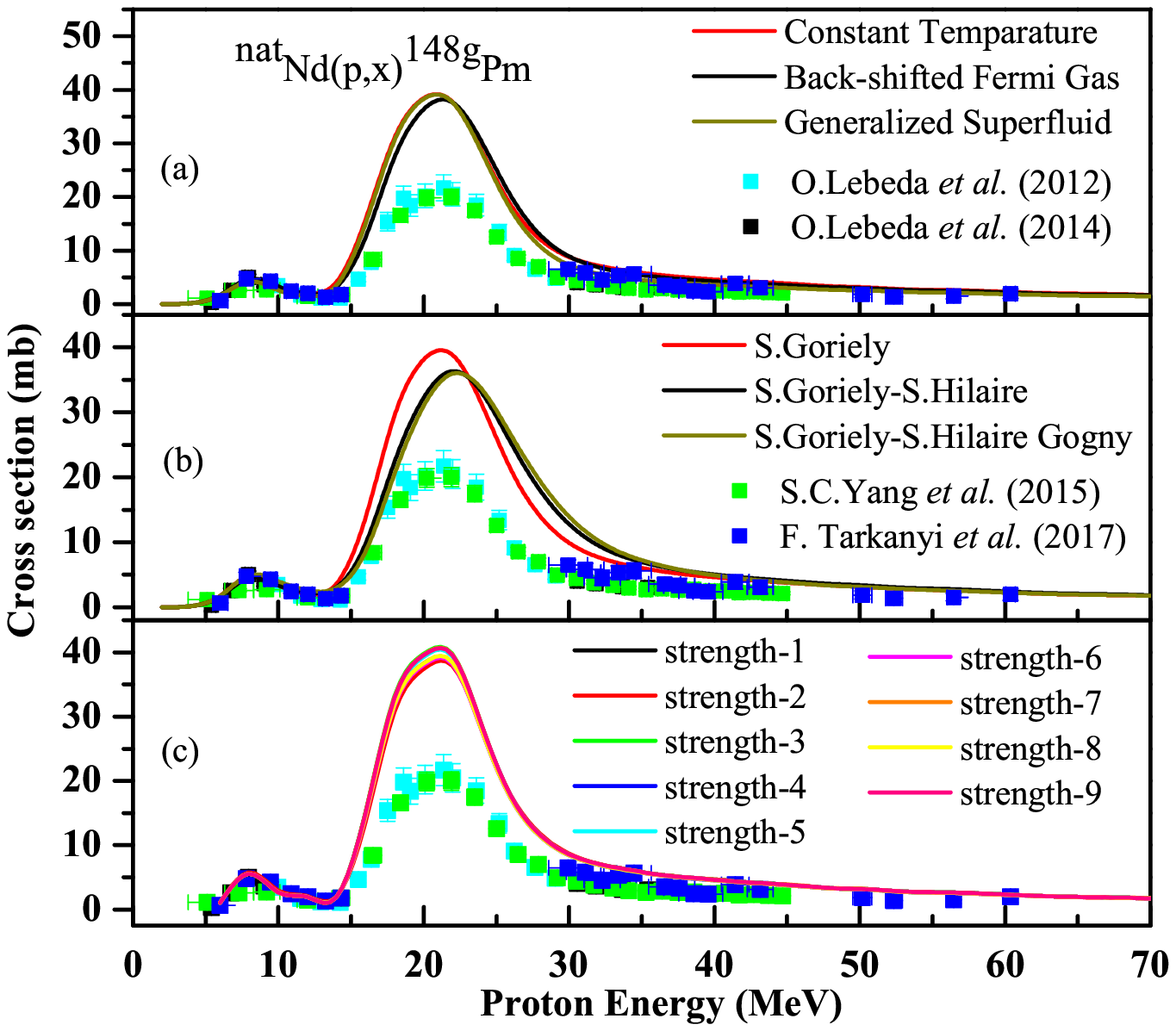}
}
\caption{Experimental data for the $^{nat}$Nd($p$,x)$^{148g}$Pm reaction reported by various groups taken from EXFOR~\cite{EXFOR} compared with the calculations using (a) phenomenological (b) microscopic level density models and (c) different gamma strength functions ($\gamma$SFs) based on the TALYS code. No significant improvement has been observed by the usage of different $\gamma$SFs when compared to the default run using different LDMs.}
\label{fig6}       
\end{figure*}

For $^{nat}$Nd($p$,x)$^{148g}$Pm reaction (see Fig.~\ref{fig6}), the predictions of the theoretical calculations based on both phenomenological and microscopic level density models matches reasonably good upto 15 MeV and beyond 30 MeV of proton energy. In the 15-30 MeV proton energy range, the various level density models when used in the default mode overestimates the cross sectional values as compared to the experimental data reported by the various groups. Also, as shown in Fig.~\ref{fig6}a and Fig.~\ref{fig6}b, slight change in the peak position of the exciation functions have been observed with the variation of the level density model.  

However, no significant improvement have been observed in the CS values for the $^{nat}$Nd($p$,x)$^{148g}$Pm reaction despite using different gamma strength functions (viz., strength-1 to strength-9) as shown in Fig.~\ref{fig6}c, in comparision to that of the default run using the six level density models.

\subsubsection{$^{nat}$Nd($p$,x)$^{146}$Pm reaction}

\begin{figure*}[!]
\resizebox{\textwidth}{!}{%
\includegraphics{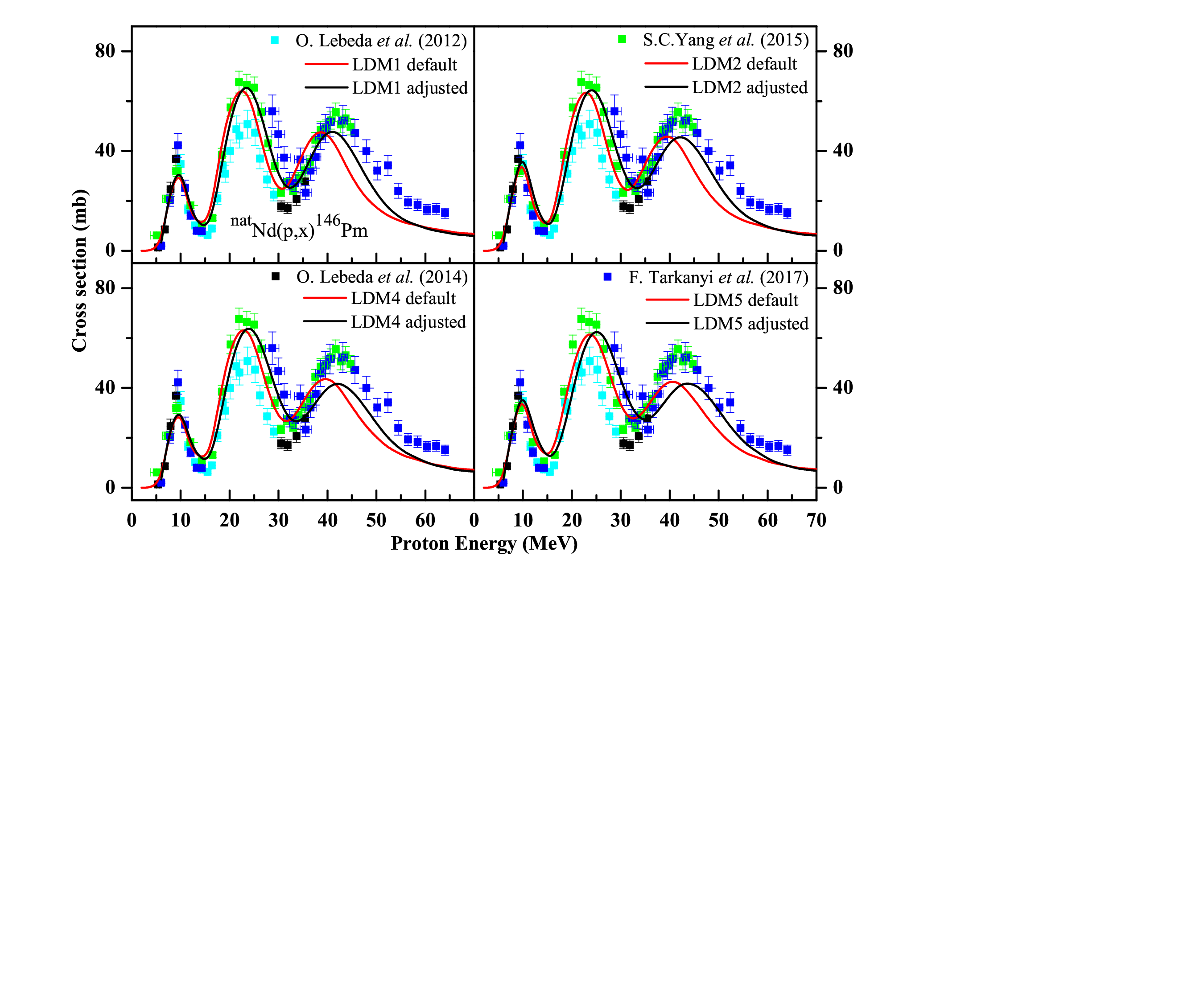}
}

\caption{The experimentally measured data for the $^{nat}$Nd($p$,x)$^{146}$Pm reaction obtained from literature compared with the phenomenological theoretical calculations based on TALYS code with different combination of model parameters. The red and the black continuous lines indicate the theoretically estimated CSs using TALYS in the default mode and using different combination of model parameters, respectively. As no significant improvement in the response have not been observed in case of LDM-3 and LDM-6 using different combination of model parameters w.r.t. the TALYS run with NLDs in the default mode, those figures have not been included here.}
\label{fig15}       
\end{figure*}

For $^{nat}$Nd($p$,x)$^{146}$Pm reaction, as shown in Fig.~\ref{fig15}, the constant temparature model and the back-shifted fermi gas model in the default mode predicts the cross section values fairly good upto 40 MeV but beyond that, these models underestimates the cross section values. The generalized superfluid model also gives satisfactory result upto 35 MeV after which mismatch between experimental results and theoretical prediction is observed. Similarly, discrepancies are observed between the theoretical predictions made by the microscopic level density models in the default mode and the experimental results beyond 40 MeV.

However, for this reaction, the response of the level density models like LDM-1, LDM-2, LDM-4 and LDM-5 improves significantly when Kopecky-Uhl generalized Lorentzian strength function (i.e., strength 1) is used along with preeqmode 3 and Bauge-Delaroche JLM potential. As no significant improvement was observed in case of LDM-3 and LDM-6 after using different combination of nuclear model parameters, the respective plots have not been included in Fig.~\ref{fig15}.

\subsubsection{$^{nat}$Nd($p$,x)$^{144,143}$Pm reaction}

\begin{figure*}

\resizebox{\textwidth}{!}{%
\includegraphics{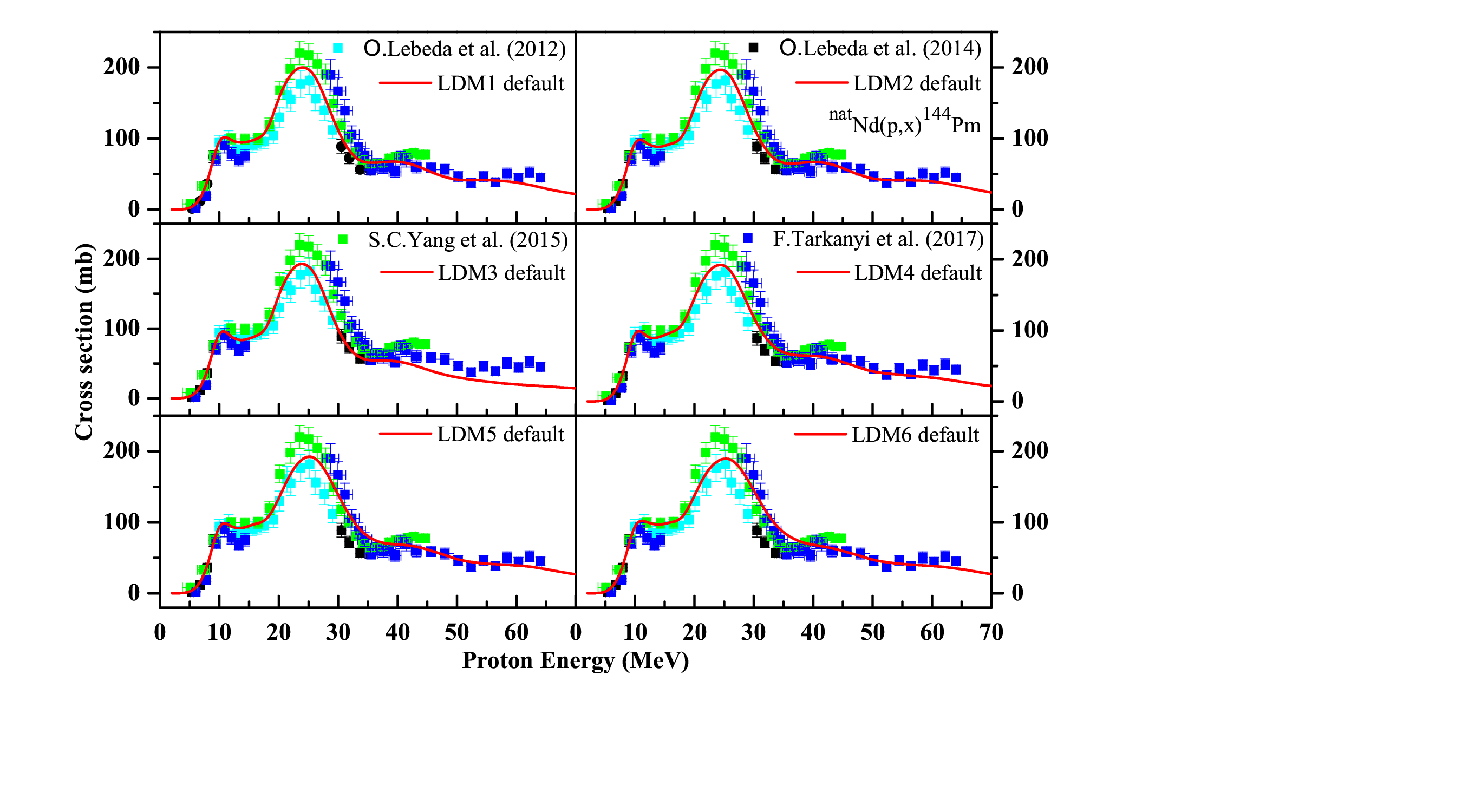}
}
\caption{Experimental data for the $^{nat}$Nd($p$,x)$^{144}$Pm reaction reported by various groups taken from EXFOR~\cite{EXFOR} compared with the six level density model calculations (i.e., LDM-1 to LDM-6) based on TALYS code with the default option.}
\label{fig8}       
\end{figure*}

\begin{figure*}

\resizebox{\textwidth}{!}{%
\includegraphics{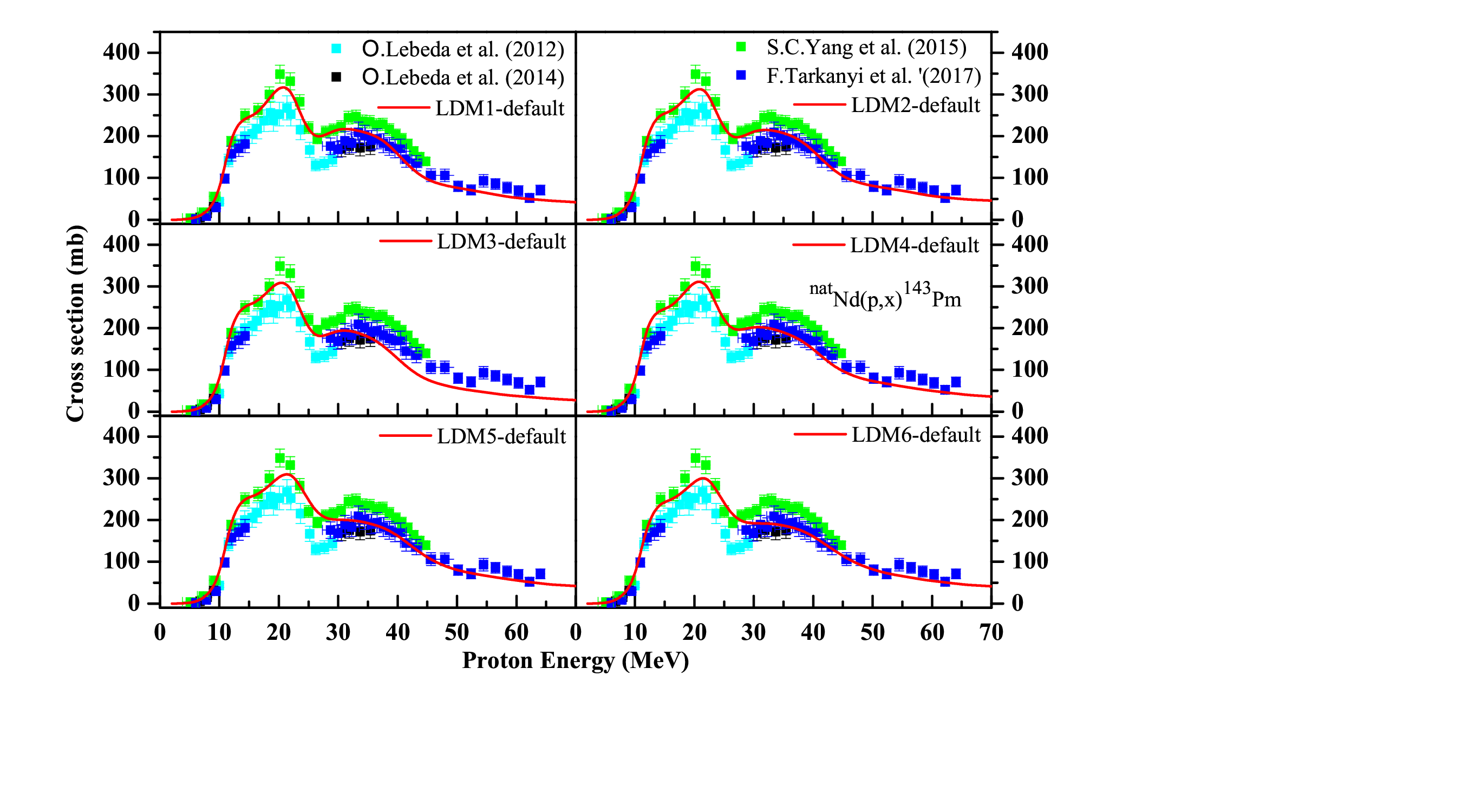}
}
\caption{Experimental data for the $^{nat}$Nd($p$,x)$^{143}$Pm reaction reported by various groups taken from EXFOR~\cite{EXFOR} compared with the six level density model calculations based on TALYS code with the default option.}
\label{fig9}       
\end{figure*}

For $^{nat}$Nd($p$,x)$^{144,143}$Pm reactions, the theoretically estimated cross section values (as shown in Fig.~\ref{fig8} and Fig.~\ref{fig9} respectively) generated using the various level density models (LDM-1 to LDM-6) in the default mode matches reasonably well with the experimental data reported by the various groups.  

\subsubsection{$^{nat}$Nd($p$,x)$^{141}$Pm reaction}

\begin{figure*}

\resizebox{0.7\textwidth}{!}{%
\includegraphics{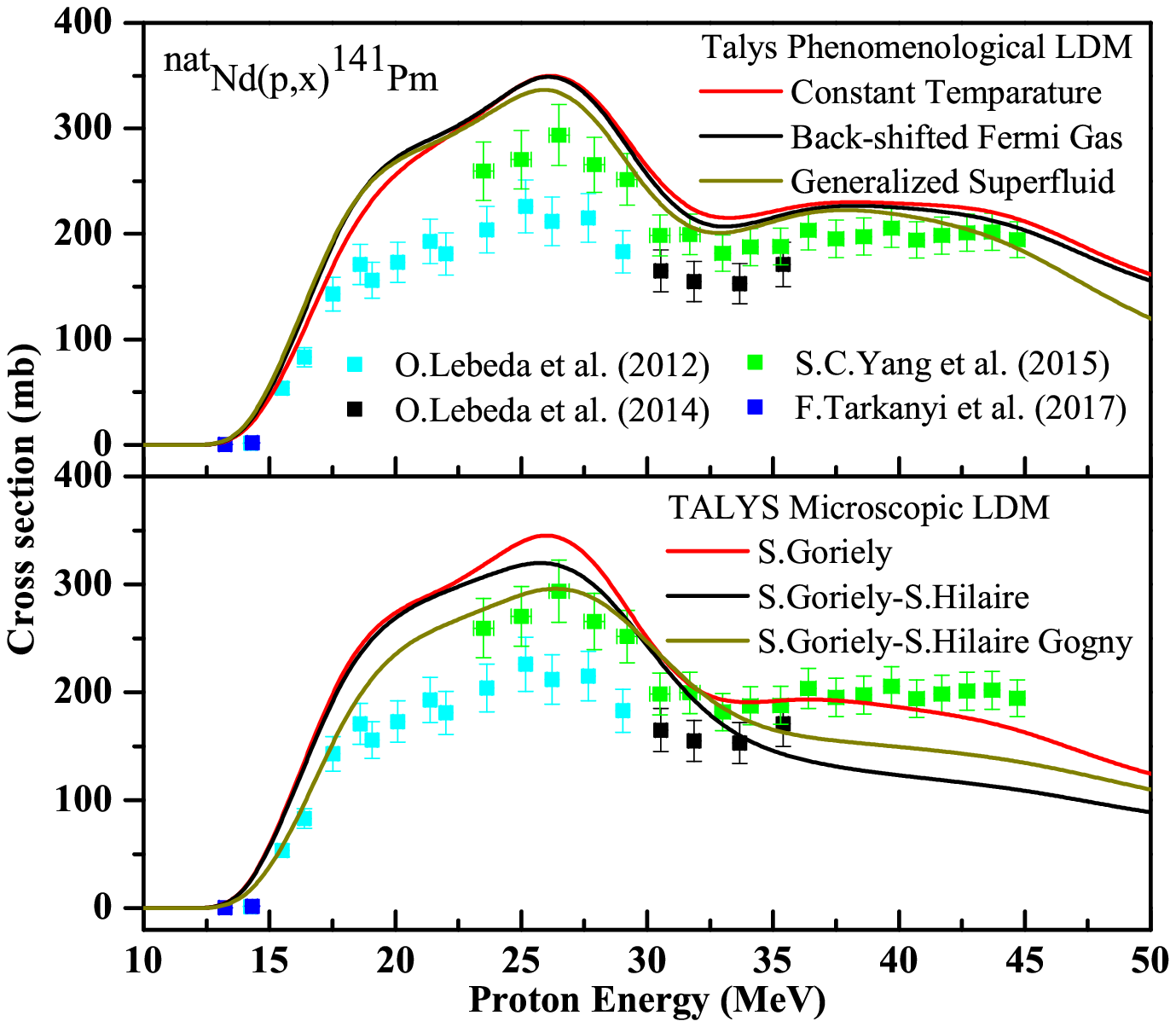}
}
\caption{Experimental data for the $^{nat}$Nd(p,x)$^{141}$Pm reaction taken from EXFOR~\cite{EXFOR} compared with the (a) phenomenological and (b) microscopic level density model calculations based on TALYS code with the default option.}
\label{fig10}       
\end{figure*}

For $^{nat}$Nd($p$,x)$^{141}$Pm reaction (see Fig.~\ref{fig10}), the phenomenological level density models in the default mode predicts the cross section values reasonably well in the 10-65 MeV range except the 20-30 MeV range where these models overestimates the cross section values. In the 35-65 MeV range, the microscopic level density models (i.e., LDM-5 and LDM-6 in particular) fail to predict the cross section values properly, when used in the default mode. 

No significant improvement have been observed in the CS values by the usage of different combination of nuclear models and model parameters as compared to that of the default run using the six level density models.

\subsubsection{$^{nat}$Nd($p$,x)$^{147}$Nd reaction}

\begin{figure*}

\resizebox{\textwidth}{!}{%
\includegraphics{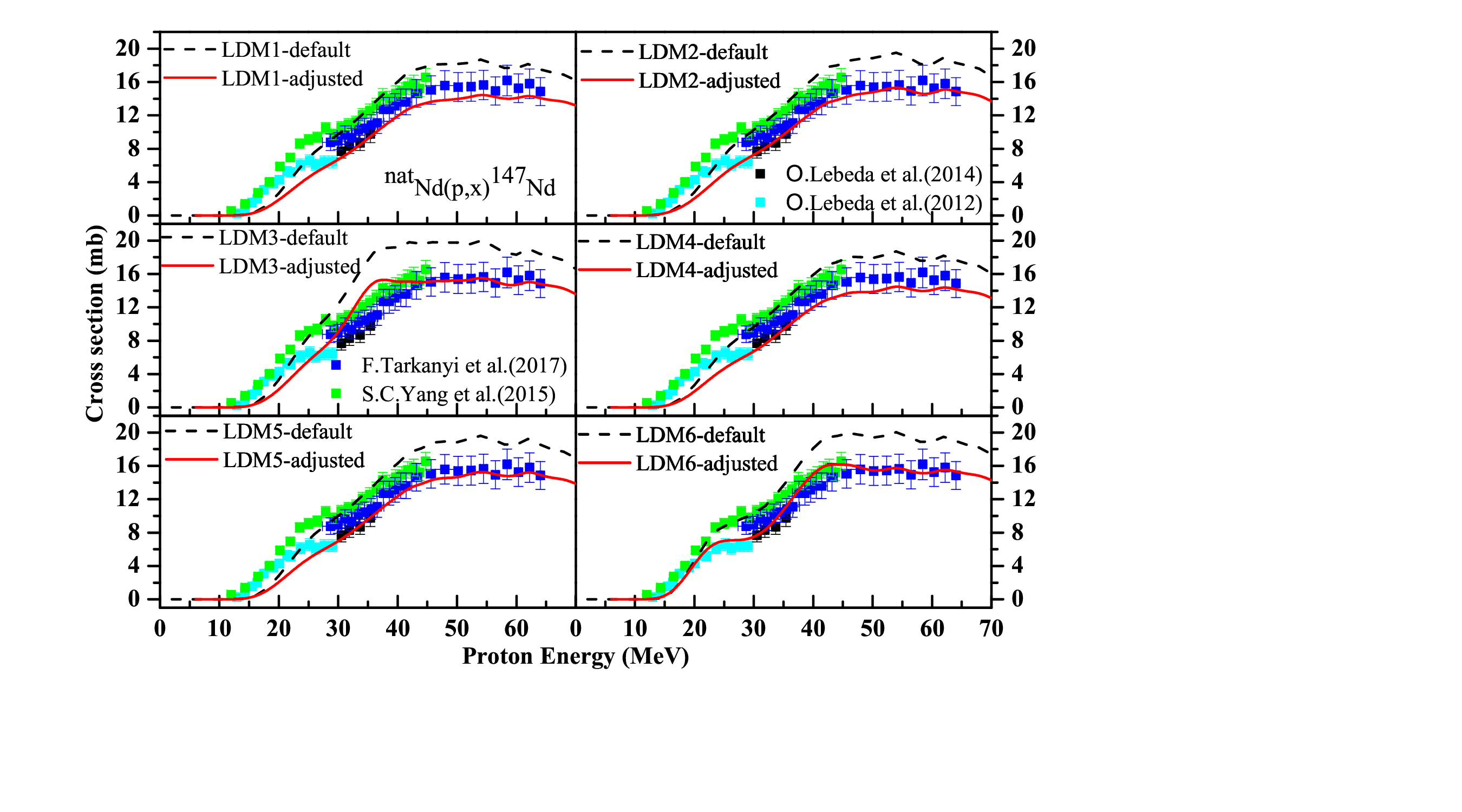}
}
\caption{The experimentally measured data for the $^{nat}$Nd($p$,x)$^{147}$Nd reaction obtained from literature compared with the phenomenological theoretical calculations based on TALYS code with different combination of model parameters. The dotted and the red continuous lines indicate the theoretically estimated CSs using TALYS in the default mode and using different combination of model parameters, respectively.}
\label{fig11}       
\end{figure*}

In case of $^{nat}$Nd($p$,x)$^{147}$Nd reaction (see Fig.~\ref{fig11}), the constant temparature model and the back-shifted fermi gas model predicts the cross section values fairly good upto 50 MeV proton beam energy after which the cross section value is slightly overestimated. The generalized superfluid model fails to predict the cross section values properly in the 30-65 MeV range, when used in the default mode. Similarly, the cross section value for the $^{nat}$Nd($p$,x)$^{147}$Nd reaction is over estimated by the microscopic level density models (i.e., LDM-5 and LDM-6 in particular) beyond 40 MeV.

However, the response of the various level density models ranging from LDM-1 to LDM-6 changes significantly for $^{150}$Nd($p$,x)$^{147}$Nd reactions when the preequilibrium model is changed from the default exciton model to the preequimode 3. Kindly refer to Fig.~\ref{fig11} to see the effect of the change of the preequilibrium mode in the response of the different level density models for the respective reaction channels.

\subsection{Comparison of experimental data with the theoretical values based on the TALYS code using adjusted OMP parameters}

{In the present work, apart from the different combinations of nuclear models and/or model parameters, an alternate process has also been adopted for fitting the various experimental data by using OMP parameters available in TALYS code. One of the motive behind this excercise was to check the sensitivity of the theoretical calculation of cross section on the OMP parameters. In order to fit the $^{150}$Nd($p$,x)$^{150}$Pm data using the TALYS code, the diffuseness parameter of the volume-central potential for the proton OMP (a$_v$) has been reduced by 20\% from its default value and the preequilibrium model has been changed from default exciton to preeqmode 3, where the collision rates are based on the optical model. For fitting the $^{150}$Nd($p$,x)$^{149}$Pm data, the diffuseness parameter a$_v$ has been reduced by 10\% from its default value and the preequilibrium model has been changed from default exciton to the multistep direct or compound model. Similarly, for $^{nat}$Nd($p$,x)$^{144,143}$Pm reactions, fine tuning of the theoretical result is done by reducing the radius parameter of the volume-central part of the proton OMP (i.e., r$_v$) value by 10\% from its default value. Similarly, for $^{nat}$Nd($p$,x)$^{141}$Pm data, the value of r$_v$ is decreased by 30\% and a$_v$ is increased by 20\%. And for $^{nat}$Nd($p$,x)$^{149,147}$Nd reactions, the values of r$_v$ and a$_v$ have been reduced by 10\% and 20\% respectively, from their default values with the purpose of fine tuning. From such systematic studies, it has been observed that adjustment of the diffuseness and radius parameter often coupled with change in the preequilibrium mode (from the default exciton mode) leads to excellent agreement between the theoretical predictions made by TALYS and the experimental data taken from the literature, in most of the aforementioned reaction channels. Please refer to Fig.~\ref{fig19} to see the effect of the aforesaid alterations in the OMP parameters (a$_v$ and/or r$_v$) and the preequilibrium mode wherever needed.

\begin{figure*}[!]
\resizebox{\textwidth}{!}{%
\includegraphics{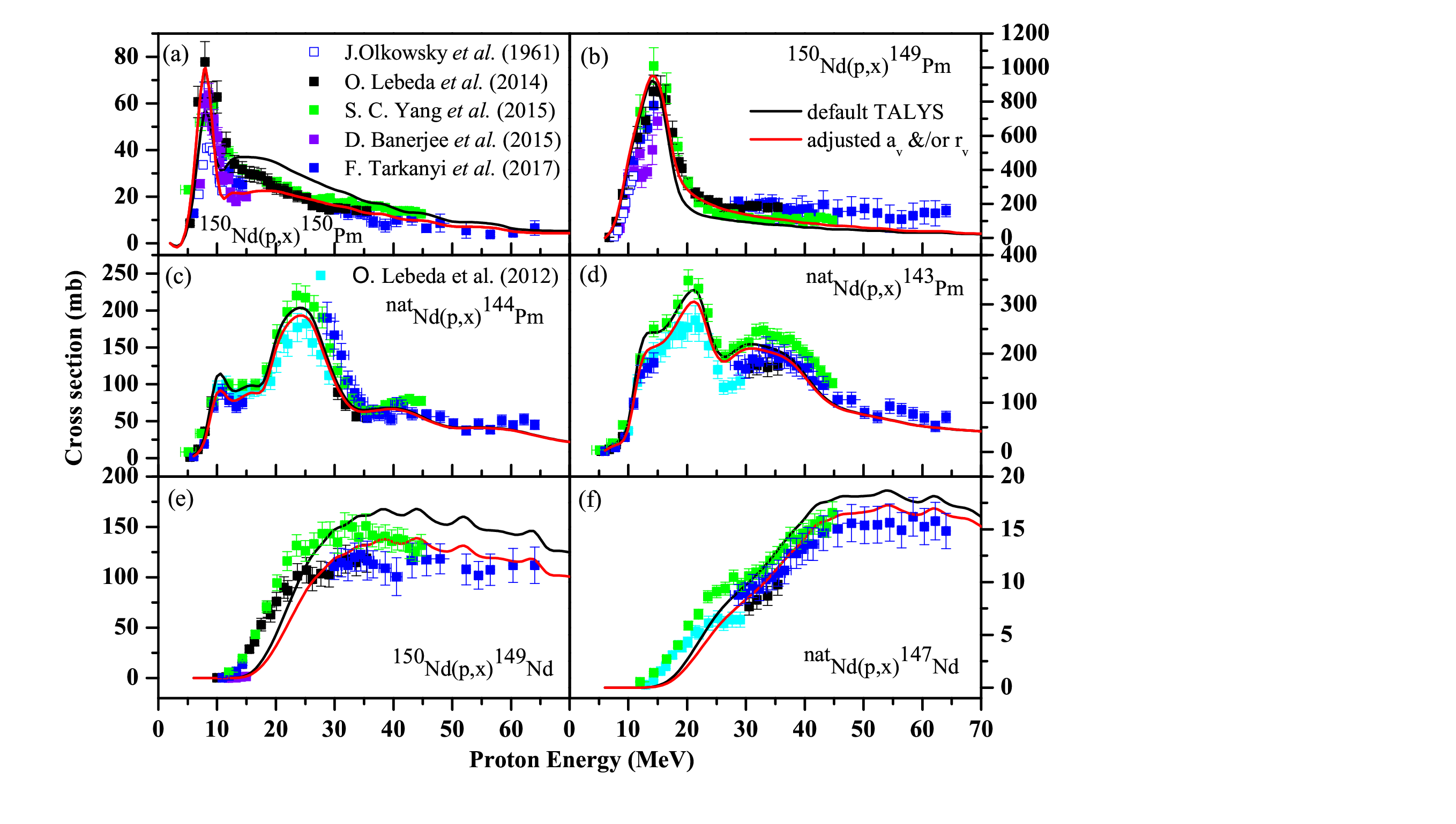}
}
\caption{The experimentally measured CS data for the $^{nat}$Nd($p$,x)$^{150,149,144,143}$Pm,$^{149,147}$Nd reactions obtained from literature compared with the default theoretical calculations based on TALYS code and with adjusted diffuseness and/or radius global parameters of volume-central part of proton OMP. The black and the red continuous lines indicate the theoretically estimated CSs using TALYS with the default mode and with adjusted OMP parameters, respectively.}
\label{fig19}       
\end{figure*}
}
\section{Summary and Conclusions}

In summary, the cross section of the $^{150}$Nd($p$,x)$^{150,149}$Pm,\\$^{149}$Nd reactions and $^{nat}$Nd($p$,x)$^{148,146,144,143,141}$Pm,$^{147}$Nd reactions were calculated theoretically using TALYS-1.96 code using both default mode as well as combination of different nuclear reaction models and model parameters available in TALYS. The cross sections estimated theoretically using TALYS code with the six level density models in the default mode were compared with the experimental cross sections available in the literature reported by the various experimental groups. Several discrepancies between theoretical estimates and experimental results were obtained when TALYS was run in its default mode. However, the various theoretically estimated cross-sections were found to be in good agreement to the experimentally observed data when certain specific nuclear parameters and models were used in combination. The impact of the various preequilibrium models and the $\gamma$-ray strength functions in the cross-section calculation is also shown in this work. Moreover, the sensitivity of the theoretically calculated cross section w.r.t the diffuseness and radius parameters of the volume-central part of the proton OMP is also investigated in the present study. It has been observed that variation of the above mentioned OMP parameters can lead to satisfactory fine tuning of the calculated cross sections for most of the reaction channels opened up via proton induced reaction on Nd target. 

The theoretically estimated cross sections for the long-lived radioisotopes like $^{143}$Pm, $^{144}$Pm, $^{146}$Pm, $^{148}$Pm, $^{147}$Nd etc. produced via proton induced reaction on $^{nat}$Nd are important for the estimation of their production rate for defining the maximum allowed time for neodymium exposure on Earth's surface, and necessary cooling down times before deploying the isotope in the SNO+ $\beta$-decay experiment.

As the range of energy values of the incident proton covered in this paper extends upto 65 MeV, it is important to mention here that the contribution of preequilibrium mechanism to the total reaction cross section attains significance at energies above 10 MeV.

\section{Acknowledgement}

The author would like to thank ICFAI University Tripura for all helps and supports for carrying out the research work. The author would also like the acknowledge the helpful discussions with Dr. Bibhabasu De and Mr. Songshaptak De.

\end{document}